# Bridging Neuroscience and AI: Environmental Enrichment as a model for forward knowledge transfer in continual learning.

Rajat Saxena[1], Bruce L. McNaughton[1,2]

1 Department of Neurobiology and Behavior, University of California, Irvine, Irvine, CA 92697, USA

2 Canadian Centre for Behavioural Neuroscience, University of Lethbridge, Lethbridge, AB, T1K 3M4 Canada

*Correspondence: brucemcn@uci.edu, rajat.saxena@uci.edu

**Acknowledgments**

The authors thank Jeffery Krichmar, Artur Luczak, Aaron Bornstein, Robert Bain, Janina Ferbinteanu, Justin Shobe, Wing Ning, and Aditi Bishnoi for their insightful comments. We also thank Aditi Bishnoi for helping us with graphic illustrations. This work was supported by NIH R01 NS121764 (BLM), NIH R01 MH125557 (BLM) and NIH RF1 NS132041 (BLM).

**Abstract**

Continual learning (CL) refers to an agent's capability to learn from a continuous stream of data and transfer knowledge without forgetting old information. One crucial aspect of CL is forward transfer, i.e., improved and faster learning on a new task by leveraging information from prior knowledge. While this ability comes naturally to biological brains, it poses a significant challenge for artificial intelligence (AI). Here, we suggest that environmental enrichment (EE) can be used as a biological model for studying forward transfer, inspiring human-like AI development. EE refers to animal studies that enhance cognitive, social, motor, and sensory stimulation and is a model for what, in humans, is referred to as 'cognitive reserve'. Enriched animals show significant improvement in learning speed and performance on new tasks, typically exhibiting forward transfer. We explore anatomical, molecular, and neuronal changes post-EE and discuss how artificial neural networks (ANNs) can be used to predict neural computation changes after enriched experiences. Finally, we provide a synergistic way of combining neuroscience and AI research that paves the path toward developing AI capable of rapid and efficient new task learning.

**Introduction**

Forward transfer is the ability to utilize previously acquired knowledge to learn new tasks rapidly and efficiently [1,2]. For example, imagine attending your first lecture on deep learning. Rather than nodding off, you're engaged, following the math, and can quickly code a simple linear deep learning model. After the class, your friend comments on how last semester's math course on linear algebra and matrices helped in today's lecture. And it suddenly hits you why the lecture felt easy. You were able to utilize and transfer concepts from math classes to the lecture on deep learning. Previous work has indicated that students with superior mathematics training perform better in computer science and logical reasoning, due to the ability to transfer abstraction skills from one subject matter to another (Fig. 1) [3,4]. The mammalian brain is capable of continual learning (CL): learning from a stream of data or experiences continuously, adapting to context changes, and realizing actions throughout its lifetime without forgetting previously acquired knowledge (catastrophic forgetting) [5]. Although seemingly trivial for humans and animals, CL remains one of the longstanding issues in AI [6,7].

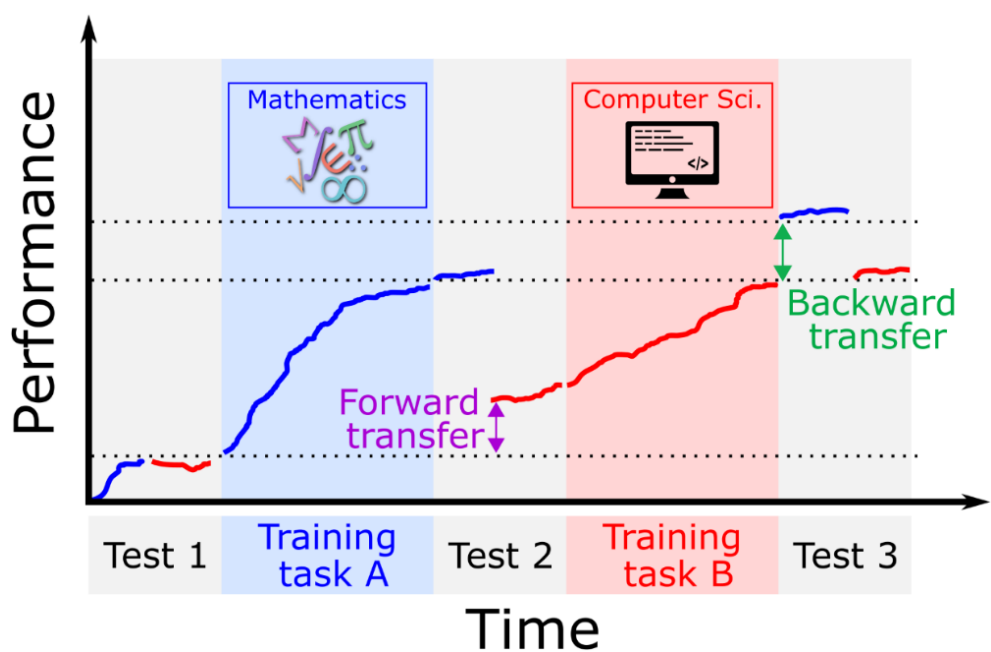

Figure 1: A schematic to demonstrate transfer learning. An agent was trained sequentially on two tasks: task A (mathematics) and task B (computer science) and evaluated on both tasks at three different times (Test 1-3, gray block), with Test 1 as the baseline. Individual task performance is expected to improve during the respective training blocks (Training tasks A and B). Due to forward transfer (purple arrow), the performance on task B (computer science, red curve) is also expected to improve in Test 2 after training on task A (mathematics). Similarly, performance on task A is *likely to improve in Test 3 after training on task B due to backward knowledge transfer (green arrow). This article concentrates solely on forward transfer (purple arrow).*

The interconnected history of AI and neuroscience, particularly cognitive neuroscience, has spurred breakthroughs in both fields. While neuroscience has continued to inform the development of brain-like algorithms (although subtly), the road to success has been challenging [7–10]. Despite surpassing expert human performance in challenging tasks such as playing Atari [11], Go [12], etc.; current ANNs struggle to capture multiple aspects of human intelligence. Most of the impressive results obtained have been in a static learning setting with a single objective, such as image classification. A CL framework is different since it involves a potentially infinite stream of data stemming from different input domains (visual, audio, etc.) with unclear task boundaries and changing task demands [6,7].

How do human brains achieve this feat of CL? Not all your friends would have been good at computer science on the first attempt. Can an individual's ability to learn a new task be influenced by their prior experiences and environment? We draw inspiration from neuroscience research examining the effects of environmental enrichment (EE), especially in rodents. EE refers to environmental modifications to provide better cognitive, social, motor, and sensory stimulation and has shown promising results in improving learning and memory [13–15]. Enrichment was first studied by Donald Hebb [16], who noticed that the rats raised in his home performed better than laboratory-raised animals in problem-solving tasks. Subsequent studies have replicated these results across species, further examining the impact of EE on learning, memory retrieval, multisensory integration, etc. EE also alters different brain mechanisms, leading to increased neurogenesis, population sparsity, long-term potentiation (LTP), etc., all of which are hypothesized to underlie efficient learning [13–15,17–19]. These studies have helped establish EE as an effective strategy for improving learning, performance, and memory retrieval - desired properties of a CL system exhibiting forward transfer. This article aims to demonstrate how EE serves as fertile ground for studying forward transfer in biological neural networks.

Amidst growing interest in developing ANNs to rival human intelligence, we propose to use EE as an inspiration for studying forward transfer in biological agents. We begin by introducing theories underlying semantic learning and the rapid acquisition of new knowledge in the brain. Then, we discuss EE and its effects on behavior and neural mechanisms. Next, we consider key insights from ANNs exhibiting forward transfer and few-shot learning (learning using very few examples). Finally, we conclude by laying down a path to combine results from EE studies and forward transfer CL in artificial agents to drive further improvement of ANNs and propose hypotheses to be tested in the brain in future experiments.

**Faster and efficient learning using schemas and compositionality**

The mammalian brain has an incredible capacity to acquire new information rapidly without forgetting pre-existing knowledge. What is the mechanism by which our brain accomplishes this? One prominent theory is the complementary learning systems theory (CLST) which states that the brain relies on complementary learning systems: hippocampus (HC) for rapid encoding of new information, while the neocortex (NC) is thought to be involved in slow learning, gradually forming structured knowledge or "schemas" from new information [20,21]. During sleep and awake rest, HC initiates new memory replay in NC, while NC spontaneously replays existing knowledge. This interleaved replay of new and old memories might allow for a gradual creation of context-independent category representations (schemas) in the NC distributed circuit (Fig. 2a). A set of such schemas reflects our generally acquired intelligence or knowledge of the world, representing a mental structure of concepts and the relation between them and can influence the new information encoding, consolidation, and retrieval [22,23]. It was proposed that we base our learning of new information on already existing, consolidated schemas, such that the degree of consistency of the new information with the existing schemas affects the rate of acquisition [22–24]. Based on this, the proposed slow learning role of NC was updated, with empirical work in both humans and animals, demonstrating that NC can act as a fast-learning system (just like HC) if new information is highly consistent with previously created schemas [23,25]. Similarly, in ANN simulations, there was an exponential speedup in learning new items with increasing consistency [26,27]. The extent of prior knowledge can differ significantly across the lifespan of an individual, starting with innate knowledge shaped primarily by evolution in childhood and transitioning to a mix of innate and learned structured knowledge driven by experience in adulthood [28]. There might be differences in the learning strategy of new items and learning speed between naïve vs. knowledge-rich brains, with the latter exhibiting faster learning (forward transfer) and reasoning-like strategy rather than a trial-and-error approach by the former. The accelerated learning in knowledge-rich agents could be due to schemas that help restrict the extent of internal representation parameter space that must be explored while learning a new task to the most relevant, task-specific, low-dimensional subspace, representing shared features and principles across tasks [29,30].

Recognizing shared features across tasks allows us to leverage prior knowledge and combine it with new information to learn more efficiently. Continuing with the example from Fig. 1, a person with a strong background in mathematics might be able to learn deep learning concepts easily, by breaking down the concepts and recognizing shared features (e.g., matrix multiplication, probability, etc.) between the two subjects (Fig. 2b). This ability to construct complex concepts and representations in novel circumstances by combining smaller, more basic reusable elements (priors) is referred to as "compositionality" [31–33]. This can allow a learning agent to generate an essentially infinite number of meaningful concepts from a finite set of basic building blocks, helping with generalizations and sample-efficient learning [34]. Despite their differences, compositionality and schemas are closely related. Schemas based on prior knowledge provide basic building blocks to generate novel and complex concepts using compositional processes. Similarly, compositionality influences how schemas develop and evolve. When we encounter new information, we can combine parts of existing to create new, more complex schemas. For example, we might combine our schemas for linear algebra, matrices, probability, etc. principles to create a new schema for deep learning (Fig. 2b). Building new ideas or schemas often relies on reusable schemas formed as a product of learning over time (and evolution) [35,36]. In conclusion, a deeper understanding of neurobiological processes underlying schema formation and compositionality might offer insights into how we learn new information quickly - a topic of interest for both neuroscience and AI.

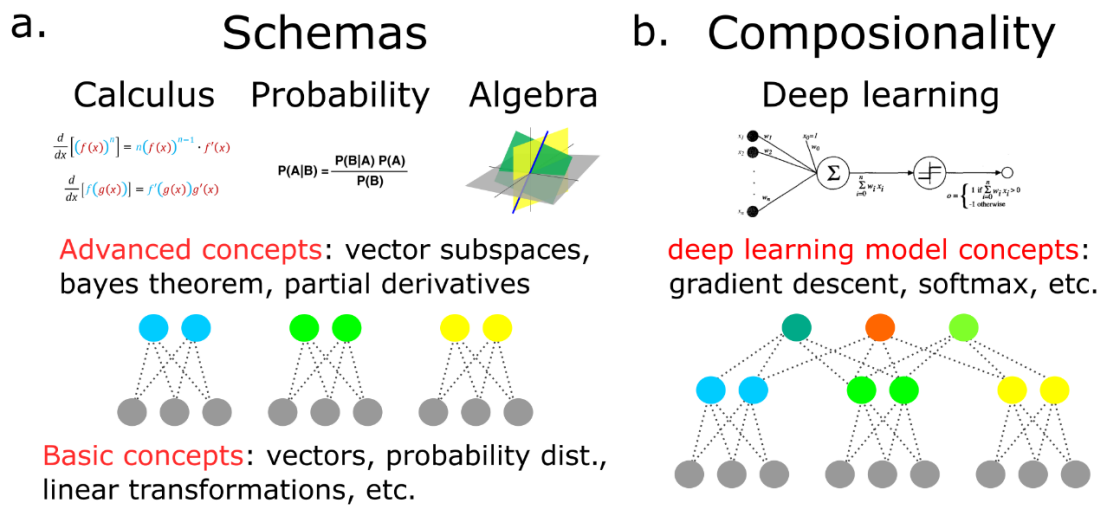

*Figure 2: a) Schemas for different math topics, including probability, linear algebra, and matrices. At the bottom are basic schemas common across topics, such as vectors, linear transformation, probability distributions, etc., that are learned first. As the individual gains knowledge of each subject matter, they develop mental abstractions (schemas) for advanced concepts (e.g., vector subspaces, partial derivatives, Bayesian probability) that are specific (and higher-order) to each topic. b) An individual with conceptual knowledge of concepts from linear algebra, probability theory, and calculus can combine them (compositionality) and learn deep learning models much faster than someone with no knowledge of those concepts. For example, stochastic gradient descent in deep learning models uses concepts of partial derivatives, matrix-vector multiplication, etc., from different math topics.*

**Environmental enrichment enhances learning, memory, and underlying brain mechanisms**

The environment around us has a significant impact on our cognitive abilities. EE is most often conceptualized as a method to improve biological and cognitive functions meeting species-specific needs, resulting from modifications to the environment, mental stimulation, social interactions, etc. [13,37,38]. Broadly, four types of EE paradigms are frequently used: physical, sensory, cognitive, and social, with each providing an opportunity to improve problem-solving skills, promote exercise, and reduce stress and anxiety or a combination of these (Fig. 3a) [14,39–41]. *Note that this article focuses mainly on cognitive EE rather than physical EE.* Apart from the type of EE paradigm, there are other vital factors of the EE paradigm that could contribute to its effectiveness. Out of these factors, age at EE onset [13,42], timing (light vs. dark phase) [43], duration [44], and continuous vs. intermittent [45] are of utmost importance [38]. For example, enrichment spanning weeks at a younger age might lead to more considerable gains than shorter exposure [42]. Many studies have tried combining different types of EE paradigms while accounting for some of the factors listed above to yield an overall more effective outcome (Fig. 3a) [38,39]. A recent study by Gattas et al. [15] used a novel EE protocol ('obstacle course') that enabled controlled enrichment delivery, accounting for physical exercise (with a control exercise track) and degree of rodent engagement with enrichment materials. The enrichment track obstacle course yielded improved and long-lasting performance on tasks involving multisensory integration, categorization, etc., as compared to standard enriched housing or increased physical exercise-only enrichment paradigm (Fig. 3a).

Extensive work has demonstrated that EE significantly improves spatial learning, spatial memory tasks, recognition memory, increased exploratory behavior, and task learning rate [14,17,40,41,45–50] (Fig. 3b, left). One possibility underlying improved behavior after EE could be the emergence of multiple schemas due to exposure to novel complex experiences, which then lead to functionally efficient connectivity, better generalization, and accelerated learning in novel situations [22–25]. Most EE studies have focused on rodent models, but similar effects have been demonstrated in humans. Physical EE (e.g., resistance training and aerobic exercise) [51], cognitive EE (e.g., number of years of education, playing strategy games) [19,52], and social EE (e.g., higher social network density) [53], have all been shown to enhance learning and memory. 'Cognitive reserve' refers to the brain's ability to maintain normal cognitive function in the face of aging or neurological diseases. Engaging in EE practices, such as education, physical exercise, social interactions, etc. providing the brain with a rich and diverse set of experiences has been hypothesized to contribute to the development and maintenance of cognitive reserve [54–57]. Altogether, EE is a promising approach that highlights the potential benefits of stimulating and varied experiences on new task learning and cognitive adaptability in both humans and animals.

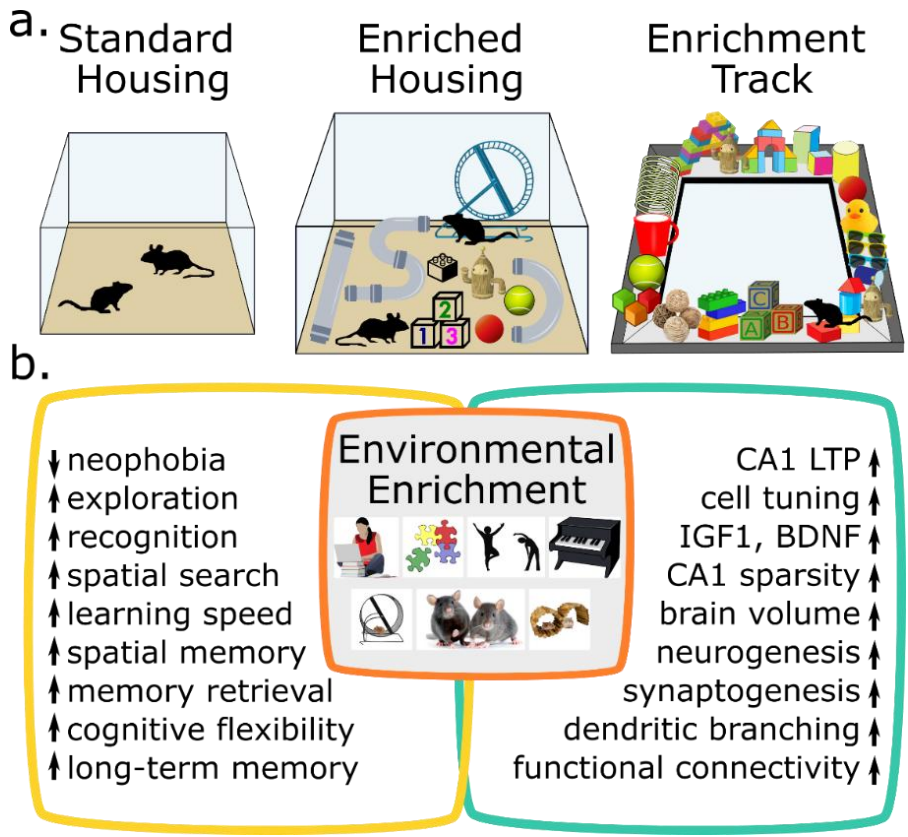

*Figure 3: a) Different types of EE strategies. Standard housing allows for no to minimal enrichment opportunities (left), enriched housing involves exposing animals to a larger arena with many objects (middle), and the enrichment track provides controlled delivery of physical and cognitive stimulation (right). b) Effects of EE (physical, cognitive, social, sensory) on behavior (left) and underlying neural dynamics (right). EE typically improves spatial memory, learning rate, exploratory behavior, recognition memory, and multisensory integration (left). EE alters mechanisms underlying learning and memory at all levels, from molecular to brain-circuit wide. Most studies report changes in increased CA1 LTP, adult neurogenesis, synaptogenesis, dendritic branching, etc. (right). Note that only a subset of effects, observed with different enrichment parameters, such as duration, types, age of initiation, etc. are illustrated.*

Along with the improvement in behaviors, EE also alters anatomical, molecular, and physiological functions (Fig. 3b, right). EE promotes synaptogenesis, and adult hippocampal neurogenesis [13,58,59], and enhances LTP in HC CA1 [60,61]. In addition, it increased spine count in basal dendrites of cortical layer II/III and apical tuft dendrites of layer V pyramidal neurons [62,63], and dendritic volume and branching [64]. Such cellular and physiological changes are associated with altered expressions of genes involved in plasticity, including increased levels of brain-derived neurotrophic factor (BDNF), N-methyl-D-aspartate (NMDA) receptor subunit 1 (NR1) [65], Insulin-like growth factor 1 (IGF-1) [66] and postsynaptic density protein 95 (PSD95) [48]. Multiple studies have also reported an increase in levels of acetylcholine, noradrenaline, and serotonin (5-HT), all indicative of triggering different learning and adaptive strategies in enriched animals [67,68]. EE also sharpens single-unit receptive fields, increases the strength of sensory cortices responses, and the selectivity, sensitivity, and efficiency of population coding (sparser coding) [69–71]. At the network level, it leads to increased macroscopic functional connectivity efficiency (higher node degree, edge connectivity) in the HC, visual, motor, retrosplenial, and cingulate cortices, which could underlie improved multisensory integration and the emergence of supramodal representations [43,72,73]. The changes discussed here may allow EE animals to adapt to new experiences and switch between different tasks, leading to improved learning, memory (forward transfer), and cognitive flexibility (Fig. 3b, right).

It is important to state here that more comprehensive experiments are needed to ascertain whether all the changes listed above contribute to improved performance and faster learning of new tasks. The results listed above were observed in EE paradigms with different designs and parameters (total duration, age of initiation, type of EE, etc.). Future studies should carefully parse out the contribution of all changes (adult neurogenesis, synaptogenesis, more integrated cortical networks, increased population sparsity, etc.) observed so far and different types of EE paradigms on schema formation to determine if they all contribute equally or if one factor outweighs the others.

**Few-shot learning and forward transfer in artificial neural networks**

EE animals clearly show aspects of forward transfer: improved performance and learning speed on new tasks [40,41]. But can ANNs exhibit similar behavior, and if they can, what factors determine those results? Take an ANN trained on linear algebra and matrices problems and teach it principal component analysis and deep learning until it becomes an expert in all. We can then ask what concepts were learned first, which were easiest to transfer, which led to maximum confusion, etc. Research on ANNs has recently shifted focus from single supervised tasks to the CL domain [7,74,75], with increased interest in joint exploration of forward transfer and catastrophic forgetting. A recent set of studies showed that ANNs with highly orthogonal prior knowledge representation show a gradient of forgetting on new task learning, with highly similar and dissimilar old tasks showing maximum and minimum forgetting, respectively [27,76,77]. For example, a network initially trained to recognize two vehicles (ship, truck) and two animals (deer, dog) showed more forgetting of other vehicle classes like planes and cars because they were similar to the original vehicle classes (Fig. 4a) [76]. To overcome the forgetting of similar classes, the model was trained to jointly replay a higher proportion of similar task exemplars (ship, truck) and novel tasks (car, plane). By exploiting the hierarchical organization of existing knowledge, this joint replay algorithm: similarity-weighted interleaved learning (SWIL), yielded rapid learning of new tasks while overcoming the catastrophic forgetting problem using substantially less data [26,27]. Additionally, the speed of learning novel items increased proportionally to the number of non-overlapping old items. SWIL also helped improve the ability of ANNs to learn new tasks from limited amounts of data and to generalize to new tasks. These results demonstrate the importance of prior knowledge in driving learning accuracy and speed, implying more forward transfer in an agent with multiple orthogonal representations [27,76,77], matching experimental work on schemas in the brain [22,23]. In the same vein, ANNs exhibit less forgetting of previously acquired knowledge and increased forward transfer as they learn multiple tasks, i.e., learning the n$^{th}$ thing gets easier than learning the first thing [1,78]. For example, learning about principal component analysis (a special case of singular value decomposition) is easier than learning about singular value decomposition for the first time. Also, relearning a previously learned task is faster than learning it for the first time (relearning savings), another characteristic of human learning and demonstrated in EE animals as well [17].

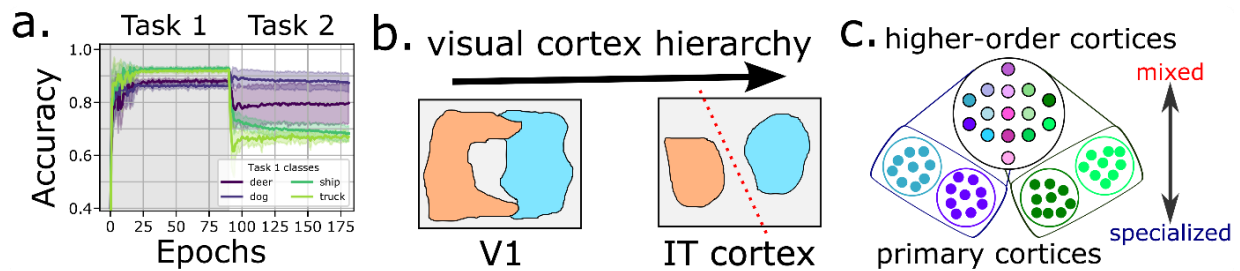

*Figure 4: a) Performance of ANN trained on two tasks simultaneously. Task 1 involved training to learn two vehicles (ship (green), truck (lime green)) and two animals (deer (indigo), dog (blue)) classes. Training the same network on Task 2 (car, plane class) leads to more forgetting of similar old classes (ship, truck). Figure modified from [76]. b) Increasing dimensionality across visual cortex hierarchy. In V1, representations for two classes (orange and sky blue) are intermixed and low-dimensional. The representation becomes high-dimensional and linearly separable in the Inferotemporal (IT) cortex. c) Bottom: representation expected in primary-sensory cortices, showing functionally specialized representations (priors) that can be reused across tasks. Top: higher-order brain regions show high-dimensional mixed selectivity abstract representations that allow for better coding, reliability, and readout by a linear classifier. Mixed-selectivity coding has been shown to provide better flexibility and linear decoding of task variables, allowing for faster and more efficient new item learning [79].*

What drives this rapid learning? Is there something different about the representations learned by the ANNs? A recent study carefully examined the relationship between the geometry of learned representations and learning speed and found that schema-like learning (prototype learning) led to the best forward transfer in ANNs on learning novel naturalistic concepts, much better compared to exemplar learning [80]. The study also reported a consistent increase in few-shot learning capability along the layers of the network, with the improvement driven by a sharp rise in the separability of class representations in the last few layers, also reported in the visual cortex [81] (Fig. 4b). Across the visual cortex (and other cortices) hierarchy, cell ensembles in individual brain regions exhibit a variety of coding schemes, ranging from functionally specialized to mixed selectivity [79,82]. This raises the question of whether there is a difference in the forward transfer capability between the two schemes and when one should be favored over the other. An elegant study attempted to address this question by training recurrent neural networks on N=20 orthogonal tasks in an interleaved design [83]. The outcome was the emergence of a cluster of units that demonstrated functional specialization and could be combined to solve novel tasks. On the other hand, the same network developed mixed selectivity after learning multiple tasks sequentially. Mixed selectivity codes have been shown to make fewer errors than highly specific codes, providing reliable decoding in sensory and high-order cortices [84]. These results indicate that pre-training ANNs on large datasets spanning a mixture of domains (akin to exposure to many experiences and tasks) might significantly improve forward transfer [80,85,86]. We propose that the higher fraction of specialized units is similar to neuronal ensembles in primary sensory cortices (more evolution-driven), allowing for reuse and sharing across tasks (compositionality), helping in tasks using a single variable. Mixed-selectivity representations, on the other hand, align well with population activity in higher-order brain regions (such as the prefrontal cortex and HC), allowing not only for compositionality and forward transfer but also simple readout by a linear classifier (Fig. 4c).

However, coding for multiple variables always raises the discrimination-generalization tradeoff concern. A modeling study reported that randomly connected neurons with a population sparsity of ~ 0.1 (10% of total neurons are active to each input stimuli) are optimal for mixed selectivity representations, as they can balance discrimination and generalization [87]. The randomly connected neurons increase the dimensionality of representations, similar to the dentate gyrus (DG) in the HC, allowing for better pattern separation [88–90]. These results hint at what might be expected in representations for a knowledge-rich vs. naïve brain. We think EE pushes the neural representations closer to a regime where generalization and discrimination (orthogonal coding) capabilities are balanced. These balanced representations might enable EE animals to learn different tasks quickly (utilizing sparse coding) and generalize across contexts. However, EE animals might show more non-linear mixed selectivity in higher-order brain regions, as it allows an efficient encoding of a large repertoire of experiences. In this section, we established that pre-training ANNs on large multi-modal datasets would help them learn generalizable representations, improving forward transfer. We pointed out that schema-like learning, exploiting the existing structured knowledge, achieves much better forward transfer compared to exemplar learning. The findings discussed here provide valuable insights into the internal representations and learning dynamics of ANNs trained to achieve rapid learning.

**Enhancing forward transfer: the neuroscience-AI nexus loop**

So far, we have looked at how EE enhances learning speed and performance on novel tasks in brains, along with results from forward transfer and few-shot learning in ANNs. Now, we want to bring the two together and discuss different mechanisms where both fields can interact and benefit from each other.

*1. Neurogenesis*:

New neurons are produced in the brain from early development to adulthood [91], most notably in DG and olfactory bulb [92]. Adult-born immature DG cells have been theorized to promote pattern separation (remapping) [93] and are essential for remote memory reconsolidation [94]. Reduced adult neurogenesis in DG extends hippocampal dependence on memory [95], indicating its role in faster integration of new information into the existing knowledge. The rate of neurogenesis is higher in EE animals, presumably underlying faster discrimination and schema formation [58,59]. Future work should focus on developing ANNs with the rate of neurogenesis altered by age and task demands. For example, lifelong learning ANNs can be trained with a larger rate of neurogenesis early on in training (similar to childhood), which can be reduced after initial deployment or have a higher neurogenesis rate throughout, identical to EE animals. Also, the higher rates could be hidden layer-specific, with the last layers exhibiting more neurogenesis than earlier layers (similar to DG) (Fig. 5a) [96]. Introducing new neurons in ANNs at different training time points could reveal its impact on categorical discrimination [97,98].

*2. Learning rule, cost function, and modular architecture*:

Most deep-learning approaches use the same learning rule and cost functions across all layers and nodes. On the other hand, learning rules and cost functions probably differ across brain regions (primary sensory vs. association cortex), and cell types, and may even change over development [99]. Also, different brain regions show different levels of plasticity and representation coding schemes, from highly specialized to mixed selectivity modules, each offering different computational benefits [79,82]. Internal representation properties of EE vs. naïve animals in different phases of learning might provide hints about the biological learning rules. We envision future ANNs to have a highly modular architecture, with each module having its own set of learning rules and cost functions (different levels of plasticity, sparsity, neuromodulators, etc.), which are configurable based on the relevance of the module to the current task (Fig. 5b) [100,101]. Based on few-shot learning ANNs (previous section), we predict that EE animals will show a more significant extent of non-linear mixed selectivity in higher-order brain regions such as the prefrontal cortex and HC than naïve animals [9,79]. Also, highly trained ANNs (knowledge-rich) tend to show a good transfer of both coarse and fine discrimination abilities, which might be true for EE animals and can be tested [28].

*3. Replay and functional connectivity:*

An interleaved replay of new and old memories during sleep has been hypothesized to improve memory consolidation and categorical knowledge [20,102]. EE animals show enhanced functional connectivity across brain regions during sleep [43,74]. A crucial difference between biological and artificial network replay is that replay in ANNs is mainly performed as a single process, whereas biological replay happens across cascaded memory systems, potentially driving changes in synaptic weights and rewiring across regions [103–107]. Future studies should add replay with different units/layers of the network involved across epochs and time (Fig. 5c). Incorporating biological network properties like sparse recurrent connectivity [108] and small-worldness (dense local clustering, short path length between distant pairs of nodes) [109], scale-freeness (connectivity degree distribution follows power law) [110], etc. might yield interesting results. Replay in ANNs could be used to understand how item representations get orthogonalized. Do coarser features separate first, followed by finer features? [8]. ANNs show larger distances between individual class representations (deeper hierarchical clustering) in the hidden layers late in training compared to the start [20]. Based on this, we predict a much more diverse exploration of network activity subspace during replay in EE animals. Also, since the distance increase is larger in the last few layers, we expect the same to be true for higher-order cortices.

*4. Neuromodulation triggered uncertainty:*

Neuromodulators such as acetylcholine, dopamine, and noradrenaline are triggered by uncertainty and unexpected rewards [101,111,112]. Previous research has hypothesized that neuromodulation assists in adapting to new experiences by initiating various forms of learning based on novelty: unknown category (create new schema) or novel exemplar in a familiar category (consolidate into an existing schema) [113,114]. Multiple studies have used neuromodulator-triggered uncertainty in ANNs to overcome catastrophic forgetting and rapid adaptation to tasks [115,116]. These attempts could significantly benefit from insights from experiments exposing EE and control animals to multiple tasks with diverse types of novelty. For example, for a new item, the network might show increased pattern separation and a larger extent of new item replay initially followed by an increased replay of previously learned knowledge [117,118]. The difference in changes in the neuromodulatory tone and its temporal dynamics across tasks in EE animals [67,68] can provide insights into novel learning rules for ANNs (Fig. 5d). ANNs can also be used to model how different uncertainty signals come into play and work together. This can guide future biological experimental designs to see if similar mechanisms are employed in brains.

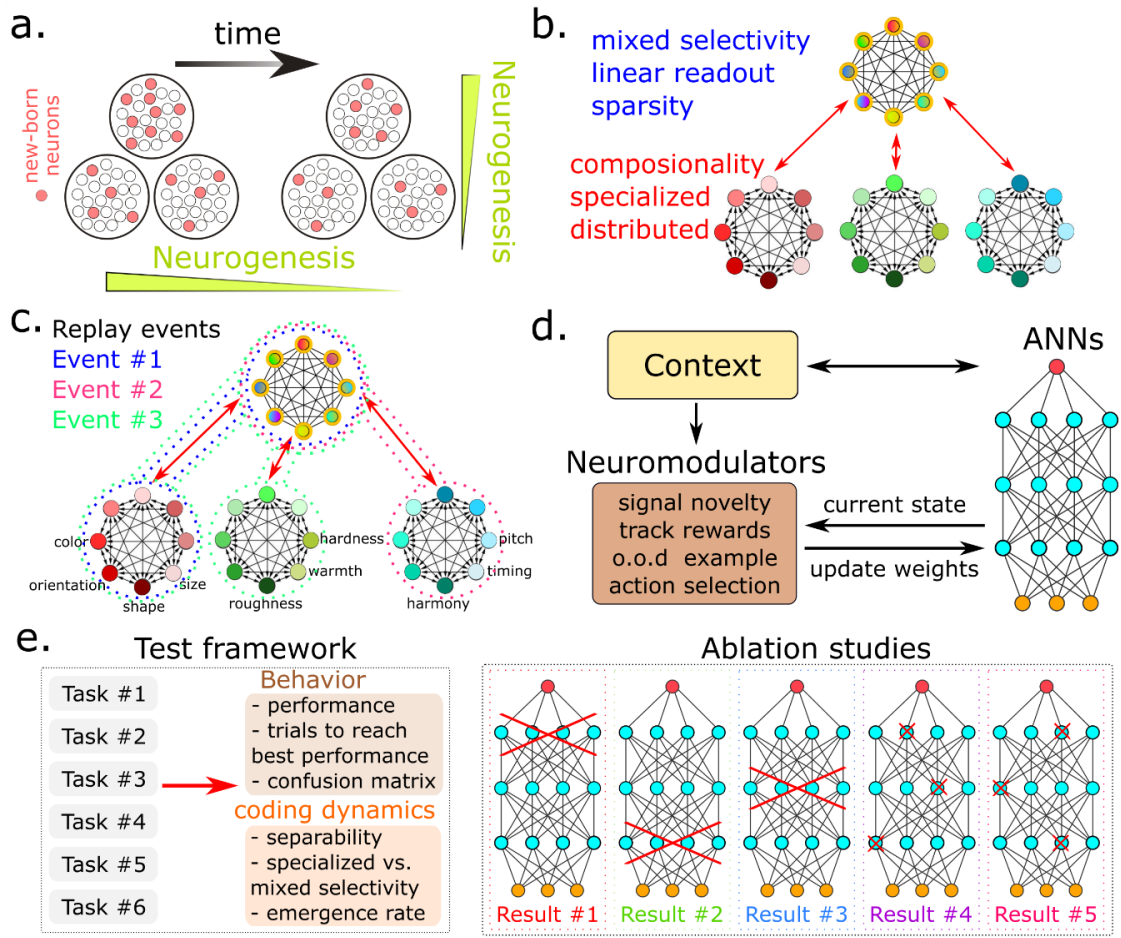

*Figure 5: Neuroscience-AI loop. a) illustration of neurogenesis in different modules of an agent across its lifespan. Note that we have shown the same number of total neurons (white + red circles) in all modules to allow better comparison concerning the fraction of newborn neurons (red circles). The agent shows a reduced rate of neurogenesis across time, from initial learning (childhood) to highly learned (adulthood) (left to right). Lower-level modules (primary sensory cortices) tend to typically have a lower rate of neurogenesis throughout than higher-level modules (hippocampus; bottom to top). b) cost functions and learning objectives are different across different modules in the brain (red vs. blue). Rather than a global rule, ANNs can be imbued with different objectives across layers and learning modules similar to our brains. c) biological memory replay involves different brain regions interacting preferentially during different replay events (blue, red, and green). A similar idea can be implemented in ANNs, where different learning modules are engaged across individual joint replay (new + old items) epochs depending on the recently learned item and gradient of forgetting. For example, replay event #1 (blue dotted line) involves replay in the top module with conjunctive coding neurons (similar to HC) and visual feature coding (red color, bottom). Replay event #3 (green dotted line) involves interaction with the top module, visual coding, and tactile coding modules (green color, bottom). We can also*
1313

*evaluate different rules for replay to see which matches the replay dynamics in biological data. Individual modules are drawn as Hopfield net in b), and c), but that is not a required condition. d) EE animals tend to show higher levels of neuromodulators and better context recognition compared to naïve animals, which might underlie rapid adaptation and forward transfer. A combination of context signal (yellow box) and neuromodulatory inputs (brown box) can be added to ANNs to update hyperparameters and make them work in different learning modes: offline memory replay phase, active learning, high discrimination, or generalization enhancing behavior performance. e) Left: Running biological experiments (Task #1- Task #6) to determine a series of parameters related to behavior and coding dynamics of a knowledge-rich vs. naïve brain. Results from these experiments can then be used to determine objective functions to enforce similar performance in ANNs. Right: Ablation (and manipulation) studies in ANNs to determine the contribution of different modules (hidden layers, first three), a subset of nodes across modules (selective for a given task, last three), and changing learning rules. Such studies can help design precise hypotheses to test in the future.*

5. *Test frameworks and ablation studies*:

We believe EE experiments can aid in developing test frameworks and datasets for ANNs, where we can come up with predictions for behavior (reaction times, time taken to achieve the best performance, number of trials required, etc.) and brain activity during learning (changes in the internal representation, increased sparsity, compositionality, etc.) (Fig. 5e, left). These results can then guide the development of rules to enable brain-like behavior performance and learning in ANNs. Future experiments can control different elements of ANNs through ablations and manipulations, which would be difficult or impossible in biological systems. This makes ANNs a powerful tool for understanding the relative contribution of each input and learning mechanism employed if some mechanisms go haywire (Fig. 5e, right) [119,120].

We haven't discussed this, but other interesting approaches such as embodiment or imparting knowledge of intuitive theories of physics can potentially help build enriched ANNs with faster knowledge acquisition and improved problem-solving [121,122]. Rapid progress in understanding novel task learning in humans and animals is imminent with the current pace of improvement in recording techniques and a steep rise in interest in neuro-inspired AI. The mechanisms and ideas listed above are just the beginning of this discussion. We believe that EE experiments and deep learning models have much to offer each other, and we hope the current paper has provided some valuable guidelines to harness the strengths of both domains.

**Concluding remarks**

In recent decades, there has been remarkable progress in behavior tracking [123] and large-scale neural recordings, going from a few hundred neurons to thousands, using Neuropixels probe [124], 2-photon mesoscope [125], etc. This advancement has enabled experiments that have provided valuable insights into existing questions and generated new hypotheses. However, just recording a large number of neurons will not be sufficient, future questions must be rooted in strong theories. Deep learning neural networks (with careful consideration) can be a powerful tool to analyze neuron representations and uncover connections between brain structure and function [126]. The aim of this article was to motivate systems neuroscience and AI researchers to interact more closely. In this article, we have discussed how EE experiments can inform new methods to enhance the forward transfer capability of ANNs. We strongly believe that combining several neuro-inspired mechanisms will significantly improve performance and speed up the development of continual learning artificial systems. Similarly, ANNs, especially deep learning models, can provide insights into the emergence of abstract representations from experience and the underlying learning dynamics. Future studies should focus on building models that can explain complex behavior and coding properties without losing interpretability to better guide neuroscience experiments. In closing, we emphasize that understanding continual learning requires a multidisciplinary effort at the intersection of psychology, neuroscience, computer science, and physics. A continuous collaboration between these fields will be critical for developing highly robust ANNs and understanding how learning and memory arise in the mammalian brain.